# Analytical correlation of routing table length index and routing path length index in hierarchical routing model

Tingrong Lu

Dept. computer science, CIDST, Beijing, 102206, China
lutingrong@hrbeu.edu.cn

**Abstract.** In Kleinrock and Kamoun's paper, the inverse relation of routing table length index and routing path length index in hierarchical routing model is illustrated. In this paper we give the analytical correlation of routing table length index and routing path length index in hierarchical routing model.

## 1 Introduction

Kleinrock and Kamoun analyzed the balance of performance parameters in hierarchical routing model [1]. According to their result, in hierarchical routing, the routing table length is reduced significantly with mild routing path length increase. Shorter routing table length results in less communication overhead. The aggregation of the routing information reduces routing table, meanwhile reduces the precision of the routing information, which makes hierarchical routing path longer than shortest routing path.

Two inversely correlated parameters, routing table length and routing path length are analyzed [1]. In the optimal hierarchical structure, the routing table length is reduced from N (the number of nodes in the network) to elnN. The upper limit of the increase of the routing path length is given.

Literature [1] supposes that the routing table length is equivalent on each node. For fixed level number, m, the optimal routing table length is $mN^{1/m}$. With variable level number, the optimal routing table length is $e \ln N$.

Though the fact of the inverse relation of the routing path length and the routing table length is well-known, to the best of the author's knowledge, there is no analytical correlation between the two parameters shown in the literature.

In this paper, we give the analytical correlation of the routing path length index and the routing table length index.

## 2 Analytical correlation of the routing path length index and the routing table length index

Theorem 1. In hierarchical structure, average routing path length stretch factor (relative length)[1], $s_p$ ($s_p \geq 1$), is linear to the number of levels of the hierarchical structure, $h$ ($h \geq 1$). $s_p = 1 + \alpha(h-1)$, in which $\alpha > 0$, is a constant decided by the rank (the number of nodes in the network) and the structure of the hierarchy.

Proof. The distance between 2 nodes equals to the sum of distance in the clusters and the distance between clusters. Assume each cluster has the same rank, the distances in the clusters are averagely the same, denoted by di. As there is only 1 hop between the adjacent clusters, the distance between 2 nodes is (1+di)*k-1, in which k is the number of clusters the path trespasses. In a tree structure, communication between any 2 nodes has to take pass on a common parent node. Denote the depth of 2 nodes, v1, v2, to their common parent node by h1, h2, the distance between v1 and v2 is d1,2=h1(di+1)+h2(di+1)-2. The diameter of the original graph, i.e. the longest distance in the graph, is 2h*(di+1)+di, in which h is the height of the tree. Assume the average path length in the original graph is d, the average path length in the hierarchical structure is d+α*(h-1), in which α>0, is a constant decided by the rank and the structure of the hierarchy. Then the average routing path length stretch factor is $(d+\alpha*(h-1))/d = 1+\alpha*(h-1)/d = 1+\beta*(h-1)$, i.e. the average routing path length stretch factor is linear to the height of the hierarchy.

From theorem 1, when h=1, the hierarchy degenerates to the original input graph, $s_p = 1+\alpha(1-1) = 1$, which satisfies the boundary condition.

Theorem 2. In IPEA model[2], the average routing path length stretch factor $s_p$ and the average routing table length stretch factor $s_t$ are in such correlation, $s_p = 1-\alpha \log s_t$ in which α>0, is a constant decided by the rank and structure of the original graph and the hierarchy.

Proof. Graph G(V, E) is devided to p clusters with same rank c, n=|V|, then $s_t = c/n = 1/p$, in which $n \geq p \geq 1, n \geq c \geq 1, s_t \leq 1$. The height of p-cluster hierarchy is $h = 1+\log p, h \geq 1$. According to theorem 1,

$$s_p = 1+\alpha*(h-1)$$
$$= 1+\alpha*(1+\log p - 1)$$
$$= 1+\alpha*\log(1/s_t)$$
$$= 1-\alpha*\log s_t$$

When the hierarchy degenerates to non-hierarchical structure, $s_t$=1, $s_p = 1-\alpha \log 1 = 1$, which satisfies the boundary condition.

In Kleinrock-Karmon model, the height of hierarchical structure is m, $m \geq 1$, optimized routing table length is $\bar{l} = mN^{1/m}$, in which N is the rank of the input graph,
$$s_t = mN^{1/m-1} \tag{1}$$
According to theorem 1, $s_p = 1+\alpha(m-1)$, we have
$$m = 1+1/\alpha*(s_p-1) \tag{2}$$
When m=1, $s_t = 1*N^{1/1-1} = 1, s_p = 1+\alpha(1-1) = 1$, which satisfies the boundary condition.

Substitute eq(1) with eq(2),
$$s_t = mN^{1/m-1}$$
$$= (1+1/\alpha*(s_p-1))N^{1/(1+1/\alpha*(s_p-1))-1} \tag{3}$$

Theorem 3. In Kleinrock-Karmon model, the average routing path length stretch factor and the average routing table length stretch factor are in such correlation,
$$s_t = mN^{1/m-1}$$
$$= (1+1/\alpha*(s_p-1))N^{1/(1+1/\alpha*(s_p-1))-1} \tag{3}$$
in which α>0, is a constant decided by the rank and the structure of the original graph and the hierarchy.

## 3   Correlation results

In Eq(3), when α=0.987, N takes 10, $10^2$, $10^3$, $10^4$, $10^5$, the respective figures are shown in Fig.1. The figures are perfectly fit to Fig.8[1].

For N=10, when $s_p$=2.2, $s_t$=0.6264, reaches the minimum. As $s_p$ increases further, $s_t$ stops decrease but begins to increase. This sounds contradictory to our empirical expectation. This is because in Kleinrock-Karmon optimal model, for N=10, $s_t$=0.6264, reaches the minimum and cannot decrease any more. In Kleinrock-Karmon optimal model, as the routing table must keep routing information of all ancestor nodes and the sub-trees of all ancestor nodes, the $s_t$ cannot decrease arbitrarily on the increase of $s_p$.

## 4   Conclusions

Eq(3) demonstrates perfectly the inverse relation between the average routing path length stretch factor and the average routing table length stretch factor in Kleinrock-Karmon hierarchical routing model, explains the data in Kleinrock-Karmon hierarchical routing model.

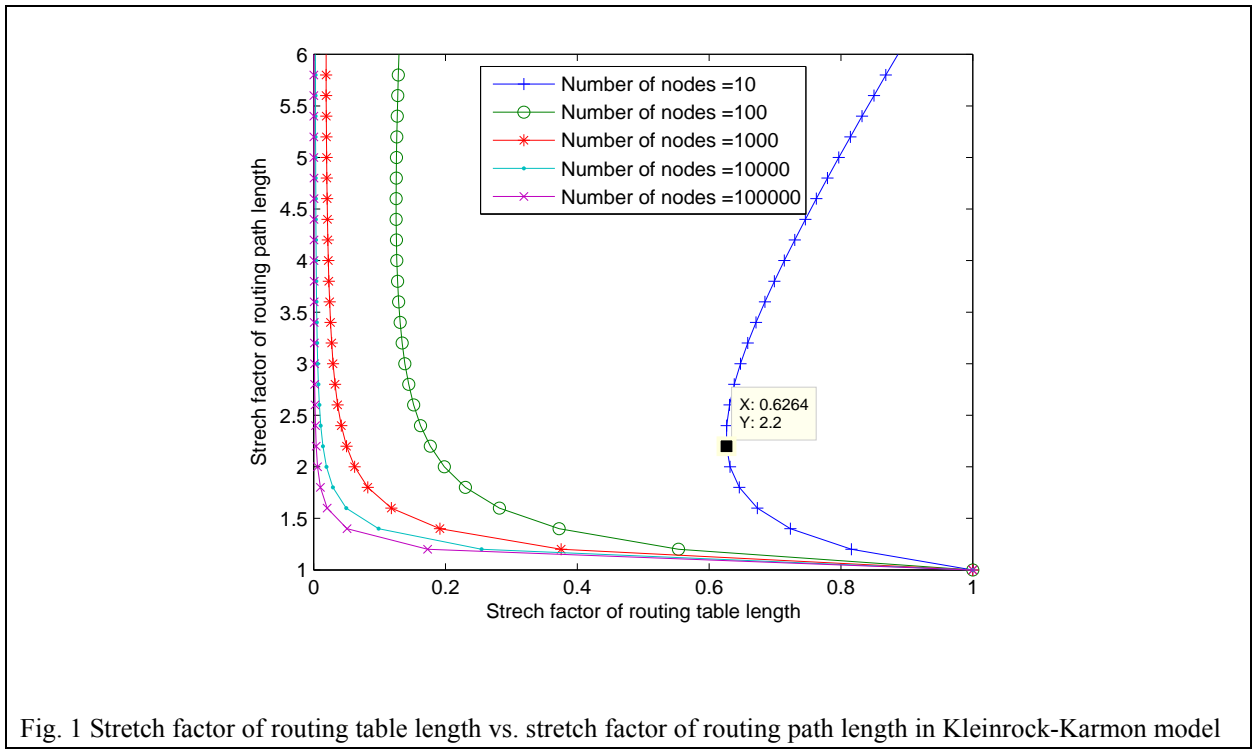

Fig. 1 Strech factor of routing table length vs. stretch factor of routing path length in Kleinrock-Karmon model